\documentclass[twocolumn,nofootinbib,superscriptaddress]{revtex4-1}
\usepackage[utf8]{inputenc}
\usepackage{longtable}
\usepackage{graphicx}
\usepackage{hyperref} 
\usepackage{placeins}
\usepackage{mathrsfs,amsmath}   
\newcommand{\sign}{\text{sign}}

\newcommand{\be}{\begin{equation}}
\newcommand{\ee}{\end{equation}}
\newcommand{\Q}{\mathcal{Q}}
\newcommand{\R}{\mathcal{R}}
\newcommand{\E}{\mathcal{E}}
\newcommand{\F}{\mathscr{F}}

\begin{document}


\title{Universal scaling and nonlinearity of aggregate price impact in financial markets}

\author{Felix Patzelt}
\email{felix@neuro.uni-bremen.de}
\affiliation{Capital Fund Management, 23 rue de l’Universit{\'e}, 75007, Paris, France}
\affiliation{Supported by Deutsche Forschungsgemeinschaft}
\author{Jean-Philippe Bouchaud}
\affiliation{Capital Fund Management, 23 rue de l’Universit{\'e}, 75007, Paris, France}
\affiliation{Ecole Polytechnique, 91120 Palaiseau, France}

\begin{abstract}
How and why stock prices move is a centuries-old question still not answered conclusively. More recently, attention shifted to higher frequencies, where trades are processed piecewise across different timescales. Here we reveal that price impact has a universal non-linear shape for trades aggregated on any intra-day scale. Its shape varies little across instruments, but drastically different master curves are obtained for order volume and -sign impact. The scaling is largely determined by the relevant Hurst exponents. We further show that extreme order flow imbalance is not associated with large returns. To the contrary, it is observed when the price is ``pinned'' to a particular level. Prices move only when there is sufficient balance in the local order flow. In fact, the probability that a trade changes the mid-price falls to zero with increasing (absolute) order-sign bias along an arc-shaped curve for all intra-day scales. Our findings challenge the widespread assumption of linear aggregate impact. They imply that market dynamics on all intra-day timescales are shaped by correlations and bilateral adaptation in the flows of liquidity provision and taking.
\end{abstract}

\maketitle

\tableofcontents


\section{Introduction}

Markets allow different sources of information to be processed and transformed into a single number: the price. Since these market prices in turn play an important signalling role for the rest of the economy, the efficiency of the price formation process is a highly relevant question. Equilibrium models explain some general features of asset prices in a formally elegant way without considering the detailed price formation process \cite{nobel2013background, farmer2009equilibrium, lyons2000microstructure}. There is, however, growing evidence that financial markets are almost never in equilibrium and that prices reflect more than fundamental information \cite{shiller1981dividends, cutler1989prices, hopman2007prices, joulin2008news}. Instead, the flow of demand and supply, information and opinions is only slowly digested, one transaction at a time \cite{Bouchaud2009MarketsSlowlyDigest}. Understanding such dynamics is of great importance for practitioners optimising their trading strategies, as well as for exchanges and regulators interested in improving market efficiency and stability.

In modern electronic markets, participants interact through a limit order book (LOB) in a continuous double-auction. Some market participants act as {\it liquidity providers} by placing limit orders (buy or sell) in the LOB. Other market participants act as {\it liquidity takers}: they need to execute their trades immediately, and correspondingly trigger transactions by sending market orders. These market orders tend to impact prices: statistically, a buy (resp. sell) market order pushes the price upwards (resp. downwards).

While the average price impact of single market orders is relatively well understood, the impact of a series of market orders is much more complex. For example, a perplexing empirical result is the square-root volume dependence of the impact of a {\it metaorder}, i.e., a sequence of individual orders belonging to the same trading decision that cannot be executed in a single transaction but must instead be fragmented (see e.g. \cite{Bouchaud2009MarketsSlowlyDigest, donier2015fully} and refs. therein)\footnote{Apparent deviations for very short or long time scales, possibly due to conditioning or undersampling, are still debated. See e.g. \cite{zarinelli2015beyond}.}. This result is at odds with, e.g., the classical Kyle model of impact which predicts a linear dependence on volume \cite{kylei1985continuous}. The empirical analysis of metaorders is difficult since it requires a proprietary database, where the trades belonging to a given trading decision can be identified. When such data is available, the square-root impact law seems to be universally vindicated, for a wide variety of markets, epochs and trading styles.

Most available datasets are, however, anonymized: while the sign $\epsilon$ and volume $v$ of each market order can be reconstructed (see section~\ref{sec:data}), the identity of the trader (or of the trading institution) at the origin of the market order is usually unknown. One can nevertheless define the \emph{aggregate impact} $\R_N$ over $N$ consecutive trades as the average price return, conditioned to a certain total volume imbalance $\Q_N$ defined as:
\be
\Q_N = \sum_{i=1}^N q_i, \qquad q_i:= \epsilon_i v_i,
\ee
where $q_i$ is the signed volume of the $i^\text{th}$ trade (see Eq.~\ref{eq:aggregate_impact} below for complete definitions). Although the impact of a single trade is well known to be a strongly concave function of its volume, it is reported that the aggregate impact of $N$ trades becomes linear in $\Q$ as $N$ increases \cite{plerou2002quantifying,Bouchaud2009MarketsSlowlyDigest}.

This picture, however, is quite incomplete as we reveal in this empirical paper. We show that once correctly rescaled, and for $N \gtrsim 10$, the aggregate impact function exhibits a non-linear, sigmoidal shape that is approximately independent of the number of transactions $N$ and of the chosen asset (large tick stocks, small tick stocks, futures). 

We also study the aggregate-sign impact, where the conditioning variable is not $\Q_N$ but rather the sign imbalance $\E_N = \sum_{i=1}^N \epsilon_i$. Scaling is again observed in this case, now with an impact function that reverts back to zero at both extremes. 

After quantifying the rescaling of the aggregate impact curves over different time-horizons, we investigate why the price-impact for extreme order-sign imbalances reverts towards zero. We find that the local bias of the order signs and the probability that an order changes the price compensate each other to a very high degree. Although possibly anticipated on general grounds, this effect does not seem to have been quantitatively reported so far, and has very fundamental and important consequences on the dynamics of markets.

The present paper is mostly about empirical observations. The ability of currently available models to describe quantitatively the scaling properties of the non-linear aggregate impact curves will be the topic of a companion paper \cite{patzelt2017nonlinear}.

\section{The Data}
\label{sec:data}

Our dataset contains the highest turnover instruments on three different platforms, namely: 
\begin{itemize}
\item 12 technology stocks on the US primary NASDAQ market, for the years 2011 to 2016. This includes some of the most traded stocks in the world like Apple (AAPL) and Microsoft (MSFT). 
\item the 13 highest turnover stocks on NASDAQ OMX NORDIC (called just OMX in the following), which covers the Nordic markets Stockholm, Helsinki, and Copenhagen for October 2011 until end of September 2015. OMX is the primary market for the selected stocks. 
\item 6 futures on EUREX EBS (BOBL, BUND, DAX, EUROSTOXX, SCHATZ, SMI) for October 2014 until the end of 2015.
\end{itemize}

We chose to analyse three different platforms in order to have some variability in terms of market microstructure in the sample while keeping the complexity of the data preparation manageable. The instruments were selected for their high turnover, reasonable concentration on their primary markets, quality of the data, and availability via the same provider for the entire period that was analysed: WOMBAT for NASDAQ, NOMURA for OMX, and the exchange itself for the EUREX data.

The NASDAQ stocks are also traded on different US-markets and trades are routed automatically to the best offer. Nevertheless, we chose to not aggregate several US-markets, because they are frequently desynchronised at the millisecond scale \cite{mackintosh2016needIII,mackintosh2016needV}, leading to inconsistent aggregate bid- and ask-prices, that is, the best visible buy and sell limit orders, respectively, as reported by the market just before each transaction is executed. We found that the microstructural parameter $\eta \in [0,1]$ appears to be a good measure of the importance of price-discretisation.%
\footnote{$\eta := N_c / (2 N_a)$, where $N_c$ is the number of subsequent price-movements in same direction (continuations) and $N_a$ the number of price-movements in alternating directions. It measures the effect of discretisation of a diffusion process. $\eta > 0.5$ corresponds to small-tick instruments and $\eta < 0.5$ to large-tick instruments. \cite{dayri2015large} \label{fn:eta}\label{footnote:eta}} %
Prices on NASDAQ are discretised with a fixed tick-size of $\$0.01$, which can be considered very small ($\eta = 0.73$) to medium ($\eta = 0.49$) for the analysed stocks. Up to roughly one third of the transactions were executed against hidden liquidity.

Stocks on OMX are only traded on one of the Nordic markets at a time, and much less fragmented than US stocks. Tick-sizes vary with price and are effectively larger ($0.24 \leq \eta \leq 0.50$) than for NASDAQ. Here, hidden liquidity represents a vanishingly small fraction of all traded volume and seems to be concentrated on the mid. Finally, the EUREX futures are not traded on other platforms at all. Tick-sizes vary considerably between moderately large ($\eta = 0.44$) and extremely large ($\eta =0.03$).

In the following, we calculate price-returns $r_t = \log m_{t+1} - \log m_t$ from the mid-prices $m$ defined as the average of the bid-price and the ask-price just before each transaction. 

We constructed order-signs by labelling all trades above the mid-price as $\epsilon=+1$ and all trades below as $\epsilon=-1$. Trades exactly at the mid-price were discarded. We decided not to use the signs provided by the exchanges themselves because hidden liquidity is not correctly labeled on NASDAQ for a part of the analysed period. Nevertheless, we confirmed all the following results using the exchange-provided signs, with only very minor quantitative differences. Trade-ids were only available from EUREX. Therefore, we merged all transactions based on the timestamps, which were reported with millisecond precision for all three platforms (see also appendix~\ref{sec:single_trade_impact}).

Trading volumes vary considerably over time. To control for extremely active days, we normalised aggregate transaction volumes $\Q = \langle Q_D \rangle / Q_D\ \sum_i q_i$ by the daily volume $Q_D$ relative to its average. This global normalisation will be omitted in the following equations for notational simplicity.

The first 30 minutes after opening and before closing on each day are discarded, as well as all days with shortened trading-hours. Obviously irregular entries were discarded too, such as transactions labelled as irregular by the exchange or provider, transactions outside the aforementioned hours, or transactions with non-finite prices (including bid- and ask-prices).

\section{Results}


\subsection{Aggregate Impact}

\begin{figure*}
  \includegraphics[width=\textwidth]{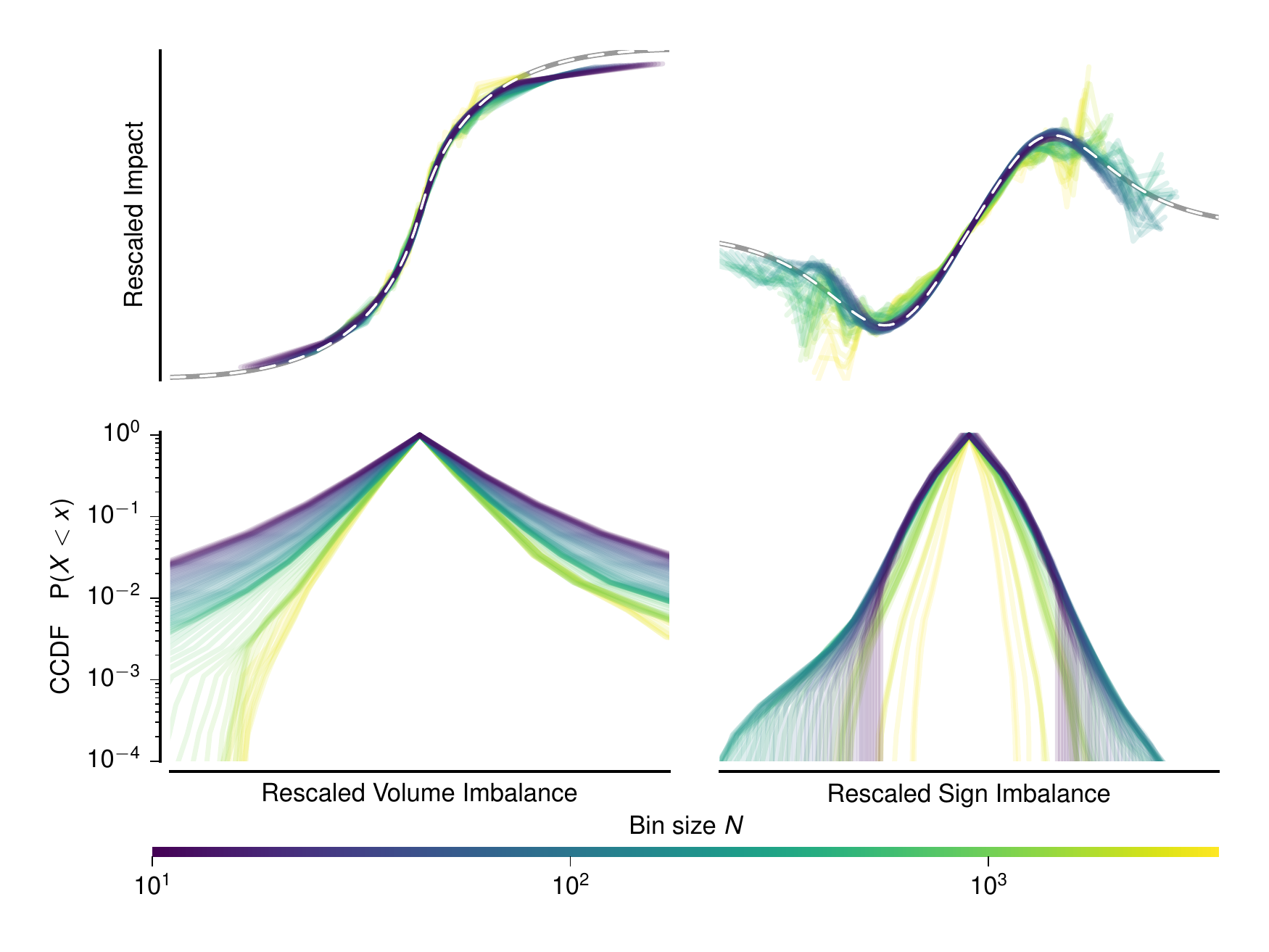}
  \caption{AAPL on NASDAQ in 2016. \textbf{Upper row, left}: rescaled expected return $\R_N(\Q / N^\xi) / N^\psi$ conditioned of the volume imbalance $\Q$ for different bin sizes $N$ in arbitrary units (see eqns.~\ref{eq:aggregate_impact}, ff.). X- and y-axis rescaling exponents: $\xi = 0.84$, $\psi = 0.53$. \textbf{Right}: rescaled mean return $\R_N(\E / N^\psi) / N^\psi$ conditioned on the sign imbalance $\E$ (see eq.~\ref{eq:aggregate_sign_impact}). $\xi_\epsilon = 0.69$, $\psi _\epsilon= 0.48$. \textbf{Lower row}: the corresponding complementary cumulative distributions. The positive and negative half were calculated independently and then binned to smooth out noise and discretisation steps for small $N$. The largest shown $N$ corresponds to the shortest day in the sample.}
  \label{fig:rescaled_detail}
\end{figure*}

As mentioned in the introduction, we measure the aggregate-volume impact as
\be
\R_N(\Q):= \left \langle \log m_{t+N} - \log m_t\, \Big{\vert}\, \Q = \sum_{i=0}^{N-1} q_{t+i} \right \rangle,
\label{eq:aggregate_impact}
\ee
where $m_t$ is the mid-price immediately before the $t^\text{th}$ transaction, $q_t$ the signed volume of the $t^\text{th}$ transaction and $\langle \dots \rangle$ denotes an empirical average over all time windows containing $N$ successive trades, executed the same day. $\R_1(\Q)$ corresponds to the average impact of a single market order of signed volume $\Q$ as studied in, e.g., \cite{lillo2003econophysics,Bouchaud2009MarketsSlowlyDigest}.

As expected, both width and height of the function $\R_N(\Q)$ increase with $N$. However, if one rescales the $\Q$-axis with an $N$-dependent volume scale $Q_N$, and the $\R$-axis with an $N$-dependent return scale $R_N$, all curves for $N \gtrsim 10$ collapse to a single master curve, as shown in Figure~\ref{fig:rescaled_detail} for AAPL, and in Appendix~\ref{sec:all_impacts} for a variety of other assets. More precisely, one finds that empirically:
\be\label{eq:scaling_form}
\R_N(\Q) \approx R_N \F\left(\frac{\Q}{Q_N}\right),
\ee
where $Q_N$ and $R_N$ both obey power-law scaling with $N$,
\begin{align}
Q_N &\approx Q_1 N^\xi,\\
R_N &\approx R_1 N^\psi,
\end{align}
and the scaling function $\F(x)$ is a sigmoidal function parameterized as
\be\label{eq:scaling_fun}
\F(x) = \frac{x}{(1 + |x|^\alpha)^{\beta/\alpha}},
\ee
where $\alpha$ and $\beta$ are fitting parameters that describe the shape of $\F(x)$. Note that for $x \to 0$, the leading behaviour is:
\be
\F(x) = x - \frac{\beta}{\alpha} \sign(x) |x|^{1+\alpha} + \dots, 
\ee
i.e., a linear behaviour with possibly non-analytic corrections.

For $x \to \infty$, on the other hand, one has:
\be
\F(x) = \sign(x) |x|^{1-\beta} + \dots. 
\ee
Hence $\beta = 1$ corresponds to saturation for large volumes, $\beta < 1$ to continued growth, and $\beta > 1$ to reversal towards lower impacts.

In order to determine the rescaling exponents $\xi$ and $\psi$, the shape of $\R_N(Q)$ is fitted for each $N$ using the scaling form Eq. (\ref{eq:scaling_form}) with 
$\F(x)$ given by Eq. (\ref{eq:scaling_fun}), keeping the same value of $\alpha$ and $\beta$ for all $N$.
\footnote{Technically, this was achieved by alternating between fitting either the scales or the shape parameters and using nonlinear regression. Only 80\% of all $N$ were randomly included in each pass.} We obtained $\alpha = 1.2 \pm 0.6$, $\beta = 1.3 \pm 0.7$ for the mean and standard deviation of the fitted $\R_N(Q)$ across all instruments in the sample. The corresponding scaling function for AAPL is shown as a dashed line in Fig.~\ref{fig:rescaled_detail}.%
\footnote{Some instruments exhibit a slight reversal of the aggregate-volume impact $\R(\Q)$ for very large arguments. These are sometimes fitted with quite large $\beta$, but the fitted curve only strongly reverts outside of the observed range of $\Q$. This is in very different from $\R(\E)$ discussed below, which strongly reverts close to zero impact within the frequently observed range of sign imbalances $\E$.}

\begin{figure}
  \centering
  \includegraphics{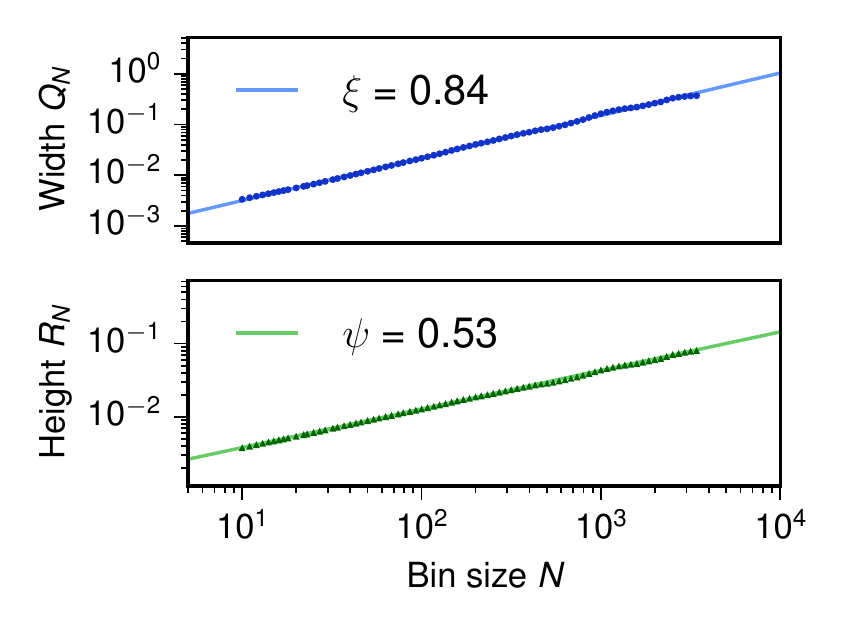}
  \caption{AAPL in 2016. \textbf{Top}: Volume scales $Q_N$ for the impact curves (see eq.~\ref{eq:scaling_form}) that were fitted to each bin-size for the same data shown in the upper left part of Fig.~\ref{fig:rescaled_detail}. Solid line: power-law fit used for the rescaling along the volume-imbalance axis. \textbf{Bottom}: The same analysis but for the return scale $R_N$, i.e. the scaling of the impact curve along the return axis. Note that the preceding fitting of the impact curves, yielding $Q_N$ and $R_N$ for each $N$, did not impose any assumptions on their scaling. Relative fitting errors are below $1\%$, but variability across instruments and periods is much larger (see below).}
  \label{fig:fit_scaling}
\end{figure}

Once $\F(x)$ is fixed, one can map out the scale factors $Q_N$ and $R_N$ as a function of $N$, which are described very accurately by power-laws of $N$ as shown in Fig.~\ref{fig:fit_scaling}.\footnote{This was done using robust regression. We also tried to fit the power-law rescaling without using parametric curves as an in-between step, but failed to achieve the same level of reliability across instruments and time periods.} The final rescaled impact functions are shown in Fig.~\ref{fig:rdV-rescaled-impacts-all} of Appendix~\ref{sec:all_impacts} for other stocks and futures. All scaling curves look remarkably similar, as indicated by the similar values of $\alpha$ and $\beta$ in all cases. Any theoretical approach will have to explain not only the value of the exponents $\xi$ and $\psi$, but also of the full master curve $\F(x)$.

Together with the rescaled aggregate-volume impact, Fig.~\ref{fig:rescaled_detail} shows the corresponding cumulative distribution of volume, rescaled by $Q_N$. Events far in the saturation regime occur with probability $\sim 10^{-2}$ on a daily basis. This must be compared with the typical number of trades per day, which is of the order of $10^4$ for AAPL. For example, there about $100$ events per day at the end of a bin of size $N=100$ and within the saturation regime. Events contributing to the saturation regime are relatively frequent, and the effect is therefore not anecdotal. 

The right-hand panel of Fig.~\ref{fig:rescaled_detail} shows the rescaled aggregate-sign impact, defined as:
\be
\R_N(\E):= \left \langle \log m_{t+N} - \log m_t\, \Big{\vert}\, \E = \sum_{i=0}^{N-1} \epsilon_{t+i} \right \rangle.
\label{eq:aggregate_sign_impact}
\ee
Here, the impact for small sign imbalances is more linear than for the volume imbalance, corresponding to a larger value of the effective parameter $\alpha$. Around a sign imbalance of $50\%$, the impact saturates sharply and reverts towards zero at the extremes. This may come as a surprise since it means that a very strong imbalance in the order-signs is associated to a very small price change on average. This effect is found for all instruments, and also for the trade imbalance, as shown in Figures~\ref{fig:rdS-rescaled-trade-impacts-all} and~\ref{fig:rdTi-rescaled-trade-impacts-all} in Appendix~\ref{sec:all_impacts}. The reason for this highly peculiar behaviour is investigated below. First, however, we have a closer look at the scaling exponents 
$\xi$ and $\psi$ (and their counterpart for the aggregate-sign impact $\xi_\epsilon$ and $\psi_\epsilon$).


\subsection{Scaling \& Hurst Exponents}

\begin{figure*}
  \centering
  \includegraphics[width=\textwidth]{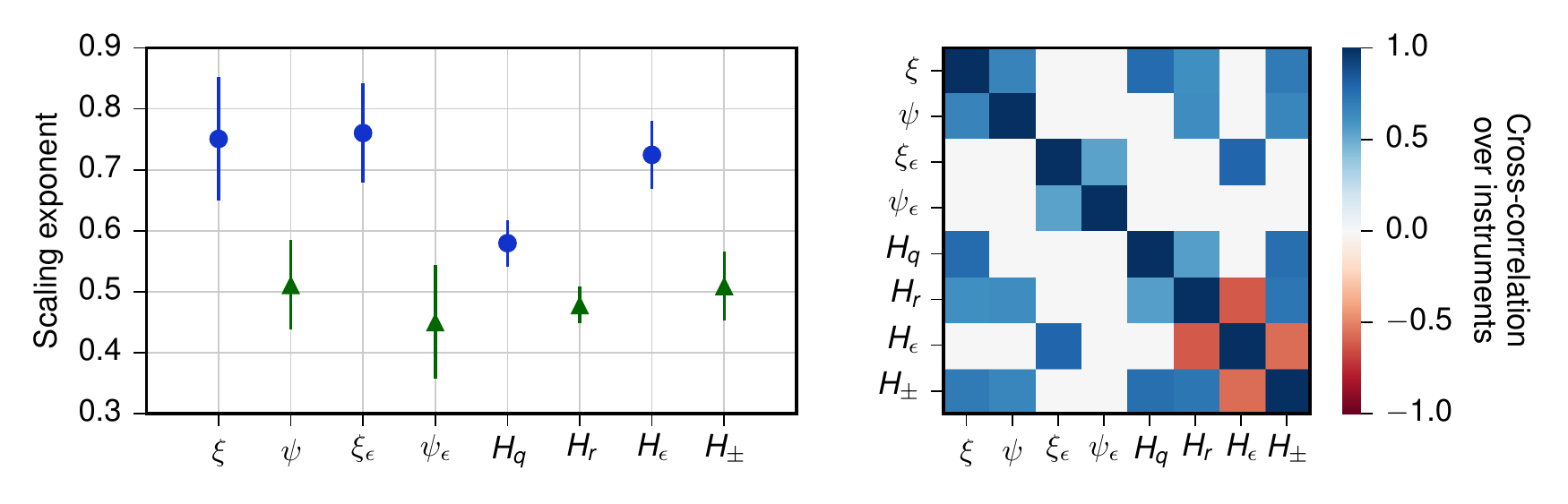}
  \caption{\textbf{Left}: Analysis of the distributions of several scaling exponents. Plotted symbols: means. Vertical lines: standard deviations. Both were calculated over instruments, for each of which the exponents were calculated for all 1-year periods before averaging. Blue dots correspond to the scaling of a variable that a price-impact is typically conditioned on, i.e. the with of an impact curve along the x-axis.  Green triangles correspond to the scaling of a variable measuring price changes, i.e. the height along the y-axis. Variables left to right: Volume imbalance and corresponding return rescaling $\xi$, $\psi$. Sign imbalance and corresponding return rescaling $\xi_\epsilon$, $\psi_\epsilon$. Hurst exponents of the Volume imbalance $H_q$, returns $H_r$, order-signs $H_\epsilon$, and return signs $H_\pm$ (see eq.~\ref{eq:hurst}).  \textbf{Right}: Cross-correlation calculated over all instruments for the same variables. Only correlations differing from zero by more then $3\,$std are shown.
}
  \label{fig:scaling_exponents}
\end{figure*}

Figure~\ref{fig:scaling_exponents} shows the means and standard deviations for several scaling exponents. The scaling exponent of the width $Q_N$ of the aggregate-volume impact is found close to $\xi \approx 0.75$ while the exponent governing the height $R_N$ is $\psi \approx 0.5$. In other words, the width of the impact curve grows faster than its height when the bin size is increased. Very similar values are found for the aggregate-sign impacts (exponents $\xi_\epsilon$ and $\psi_\epsilon$). Note that using Eqs. (\ref{eq:scaling_form}) and (\ref{eq:scaling_fun}), the slope of the linear region of impact follows as $\partial\, \R_N(\Q) / \partial\, \Q |_{\Q=0} = R_N / Q_N$. It scales as $N^{-\kappa}$ with $\kappa =\xi -\psi \approx {0.25}$, i.e. it decreases with $N$ as a power-law $N$.\footnote{The cross-instrument dispersion of $\kappa$ around its average $\langle \kappa \rangle = 0.23$ is actually relatively small: $\text{std}(\kappa) = 0.10$.}

Since both $\R_N$ and $\Q_N$ are sums over random variables (resp. returns and signed volumes), one expects that their natural scaling with $N$ is governed by the Hurst exponents of the underlying variables $r$ and $q$. Here we define the Hurst exponent $H$ of a zero-mean random variable $x$ from the scaling of the standard deviation of sums of $N$ successive events:
\begin{equation}
	\label{eq:hurst}
	\left\langle \left(\sum_{i=1}^{N} x_i\right)^2 \right\rangle := D\, N^{2H_x},
\end{equation}
where $D$ is a constant and where we take advantage of the fact that returns and signed volumes have a zero average over long time-horizons.\footnote{This method turned out to be more robust than the standard rescaled range analysis. The reason is that returns and volumes follow very heavy-tailed distributions, to which the range is much more sensitive than the sum. It is also simpler than detrended fluctuation analysis, which did not seem to be beneficial in this particular use case.}
As usual, $H=0.5$ corresponds to regular diffusion, $H < 0.5$ to sudiffusion, and $H > 0.5$ to superdiffusion.

Returns are almost diffusive with a very slight tendency for mean-reversion (particularly for large-tick instruments). This is consistent with $\psi$ and $\psi_\epsilon$. Volume signs are only slightly positively correlated, at least when measured through $H_q$. Order signs are (as is well known) strongly correlated with $H_\epsilon > 0.7$, close to the values of $\xi$ and $\xi_\epsilon$. This implies that the scales of the aggregate impact curves are mostly determined by the accumulated variation in the return- and sign- time series. Interestingly, the scaling of the impact curves is not trivially related to the Hurst exponent of volume fluctuations, but rather to sign fluctuations. This is expected from the fact that the volume of individual orders exhibits extreme variability that mostly reflects the available liquidity at the best price \cite{jones1994transactions, farmer2004really, gomber2015liquidity}. Large fluctuations of order volumes $v$ introduce an independent source of noise that masks part of the order-sign correlations when measuring $H_q$ in the way described above (see \cite{bouchaud2006random} for a related discussion). 

On the right-hand panel in Fig.~\ref{fig:scaling_exponents}, significant cross correlations across instruments between the different scaling exponents are shown.%
\footnote{Note that the square of the cross correlation can also be interpreted as the coefficient of determination $R^2$ of a linear regression with intercept.} %
We find positive correlations between $\xi$ and $H_q$ as well as between $\psi$ and $H_r$, offering some reassurance that the similar average values (left-hand pane) are more than a pure coincidence. We also find positive correlations between $\xi_\epsilon$ and $H_\epsilon$.  However, there are no significant correlations between $\psi_\epsilon$ and $H_r$, or between $H_\epsilon$ and $H_q$. This hints at a quite complex interplay of several factors driving the variability in the different scaling behaviors across instruments. $H_\epsilon$ is negatively correlated with $H_r$, which implies more order-splitting on more mean-reverting, larger tick instruments. Remember, though, that $H_r$ only varies very little across instruments (Fig. ~\ref{fig:scaling_exponents}).%
\footnote{We found consistent results measuring order-sign autocorrelations directly (not shown). The latter decay with an exponent $\gamma \approx 0.5$ for long lags. They often exhibit a steeper initial decay, however. Correlations for $\gamma$ and other exponents are similar to $H_\epsilon$, but generally weaker even though $\gamma$ varies more across assets.}%
$^,$%
\footnote{We also found that the ``implied spread'' $\eta$ (see \cite{dayri2015large} and footnote \ref{footnote:eta}) seems to be informative (not shown). It varies considerably across instruments (see section \ref{sec:data}) and is strongly correlated with several exponents. It shares these correlations with the return-sign Hurst-exponent. This finding seems to be related to the co-occurrence of larger $H_\pm$ and a larger fraction of trend continuations $N_c$.}


\begin{figure*}
  \centering
  \includegraphics[width=\textwidth]{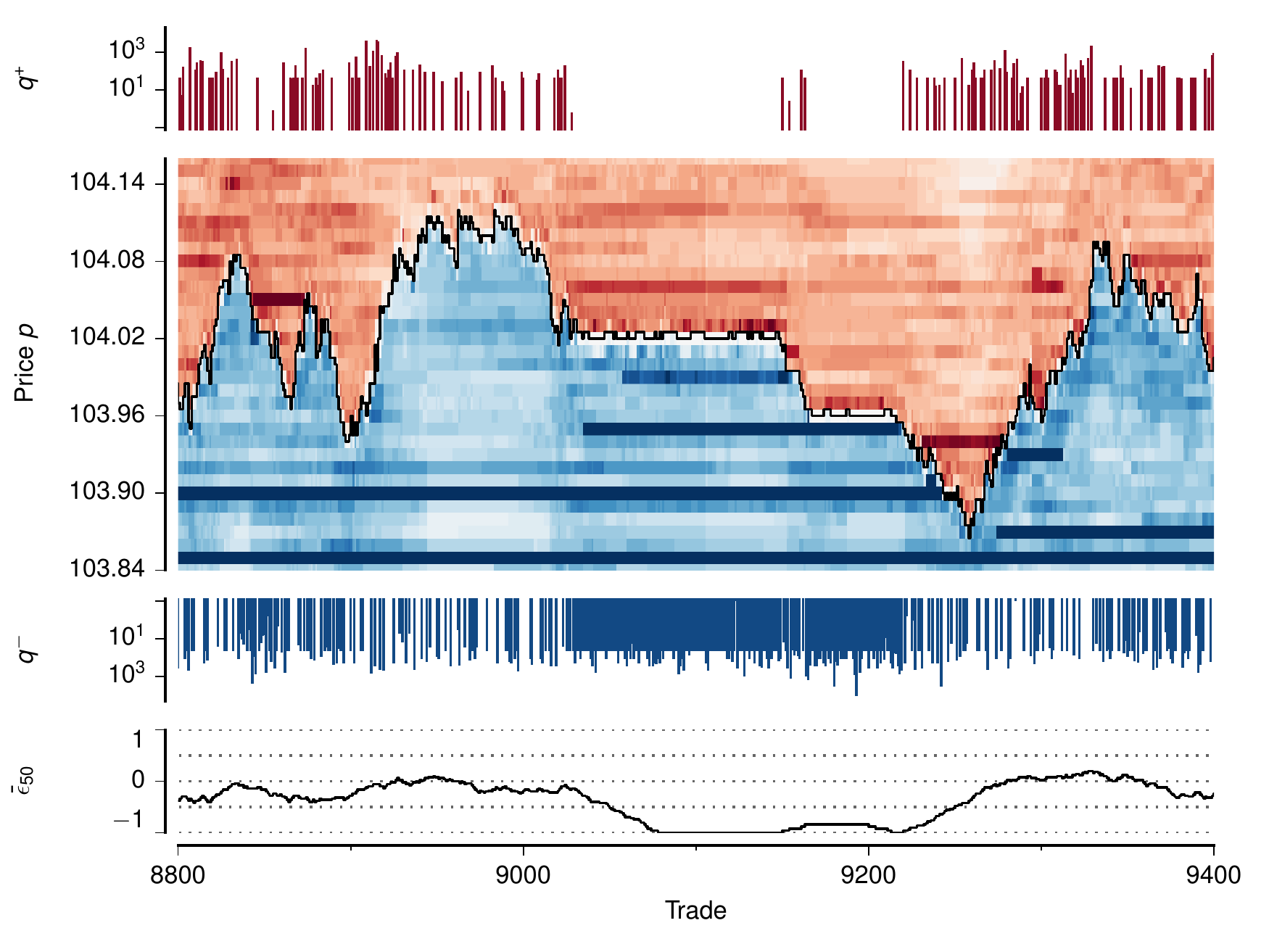}
  \caption{AAPL, May first 2016. An typical example of an extreme order-sign imbalance and a ``sticky price''. The visible liquidity at each price-level just before each trade is drawn in red above and in blue below the mid-price (upper solid line). Full saturation corresponds to the $90$-percentile for the day. The signed buy volumes $q^+$ for each trade are drawn as bars above the orderbook reconstruction and the signed sell volumes $q^-$ below. The volume axis is log-scaled. A lower bound for the true number of trades is shown because transactions with the same sign and timestamp were merged (see section \ref{sec:data}). Therefore, a series of 100 successive market buys drawn above took at least $100$ms in wall time and may have consisted of more than 100 independent market orders submitted to the market. Lower solid curve: 50 trade sign imbalance (causal). }
  \label{fig:pinned_book_AAPL_1}
\end{figure*}

\subsection{Large Order Imbalances \& Pinned Prices}

As stated before, extreme order-sign imbalance is not associated with large returns. To the contrary, such large imbalances are observed when the price is ``pinned'' to a particular level. A typical example is shown in Fig.~\ref{fig:pinned_book_AAPL_1}. Around trade $9100$, a series of more than 100 market sell orders arrives, yet the mid-price only bounces up and down half a tick as liquidity providers replenish limit buy orders at the best price or inside of the spread (the distance between the best bid- and the ask price). Many of these trades are quite small, but some are significantly larger as liquidity takers adapt to the available volume at the best. 

\begin{figure}
  \centering
  \includegraphics{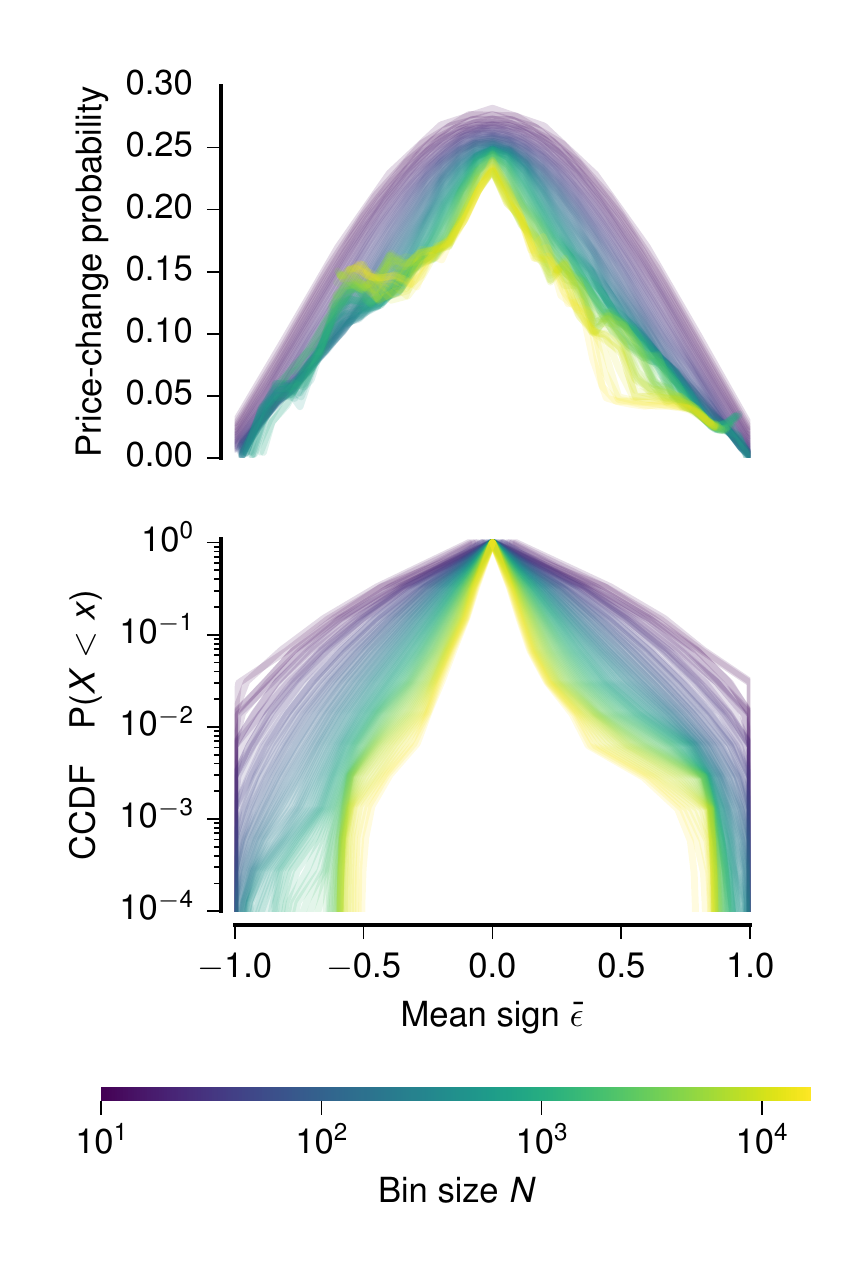}
  \caption{EUROSTOXX on EUREX in 2015. \textbf{Upper row}: the probability $P(\R \neq 0 | N^{-1}\E)$ that the mid-price changes from one trade to the next conditioned on the mean trade sign in the bin containing $N$ trades. \textbf{Lower row}: the corresponding complementary cumulative distributions.}
  \label{fig:change_detail_EUROSTOXX}
\end{figure}

The mid-price only diffuses freely when market trades happen on both sides of the book. The upper panel in Fig.~\ref{fig:change_detail_EUROSTOXX} shows the probability of a mid-price change between two subsequent trades as a function of the average order sign in the bin, $\bar \epsilon:= \E / N$. One of the central results of this paper is that biased order-signs lead to a {\it lower} probability for price changes on any intra-day time scale. Note in particular that the price-change probability has an almost invariant triangular shape for all bin sizes of $N \gtrsim 50$. It approaches zero for highly biased order signs. The qualitative behaviour is the same for smaller $N$, although the curve is broader and smoother, and has a flatter maximum around $\bar \epsilon = 0$. This behaviour is universal across all instruments (see Appendix~\ref{sec:change-all} for more examples). The corresponding cumulative distribution is shown in the lower pane, confirming that strong order-sign imbalances happen relatively frequently on a daily basis. The non-intuitive negative correlation between sign imbalance and return is therefore not an artefact due to a lack of data.  

\section{Discussion}

We investigated how prices are impacted by the flow of market orders and found a universal behaviour on all intra-day time scales. We have shown that the impact curves, once correctly rescaled, are remarkably stable across time scales (from bins of $N=10$ trades up to an entire day of trading, beyond which overnight effects would have to be taken into a account) and instruments (large \& small tick US stocks, Nordic stocks, EUREX futures). This illustrates how measuring master curves instead of either scaling laws for scalar quantities or conditional expectation curves on individual timescales can substantially improve the insights gained on market dynamics. To fully appreciate the  master curves' robustness, we kindly encourage the reader to read the appendices below. Our results suggest that the price formation process in financial markets is the result of some general, universal mechanism. While the latter is still to be elicited, our findings do provide some hints.

We find that the aggregate-volume impact saturates for large (rescaled) imbalances, on all time scales. The behaviour of the aggregate-sign impact is even more striking: highly biased order flows are associated with very small price changes.  More precisely, we find that the probability for an order to change the price decreases with the local imbalance, and vanishes when the order signs are locally strongly biased in one direction. At high frequencies, extreme order-sign imbalances occur when a very large volume is available on the opposite side of the order book, resulting in prices being temporarily pinned to a certain level. These large volumes manifest themselves either as visible large limit orders or as repeated refills near a particular price-level.

Qualitatively, this dependence of the price-change probabilities on average order signs is consistent with models and empirical results in the literature. For example, the Madhavan-Richardson-Roomans model \cite{madhavan1997security} postulates that the change of price is proportional to the sign ``surprise'', i.e. the difference between the sign $\epsilon$ of a market order and its expected value, based on previous signs. When it is extremely likely that the next trade is a buy, and a buy trade indeed materializes, then the price change is small. Empirically, many studies have reported that returns in the direction of a particular trade-sign predictor are on average lower than those in the opposite direction (see e.g. \cite{taranto2014adaptive, lillo2004long, gerig2008theory}). This effect is attributed to liquidity takers adjusting their market order volume at the outstanding liquidity, while liquidity providers revise their limit orders and refill to match the incoming order bias. Our results show a connection to the aggregate price-impact and a direct measure of the aforementioned, hypothesised bilateral order-flow adaptation. The probability of price changes as a function of local sign imbalance becomes (for large $N$) a tent-shaped function that has a discontinuous slope for zero imbalance and vanishes for strong imbalances. This observation is consistent with \cite{plerou2002quantifying} where a similar shape was reported for the standard deviation of price-changes in 15 minute windows for US stocks traded in 1994--1995, i.e. before electronic markets and High-Frequency algorithms.

Taken together, our findings suggest that markets generally operate in a state where traders collectively counterbalance the impact of predictable events to a very high degree, and on all intra-day timescales. Since market-order volumes are known to be highly conditioned on visible liquidity (see e.g. \cite{gomber2015liquidity}), the dependence of market order signs on repeated refills (as shown in Fig.~\ref{fig:pinned_book_AAPL_1}) should not come as a surprise: these observation simply confirm that liquidity takers pay attention to the currently available liquidity. Reciprocally, liquidity providers observe the flow of market orders and adapt their behavior to the well-known long-range correlations of order-signs (see \cite{bouchaud2006random} for a related discussion).

In this scenario, price fluctuations mostly reflect the a lack of predictability, or ``surprise'' of an event. Dynamics of this type have previously been shown to be capable of generating clustered volatility and extreme price-jumps in stylised multi-agent systems \cite{patzelt2013instability} and in highly adaptive control systems in \cite{patzelt2011critical}. Therefore, the present work provides a first step towards more directly testable models along these lines, and suggests that the classical notion of market efficiency \cite{fama1970efficientMarkets} should be extended to include endogenous information on top of exogeneous news.

In a forthcoming paper \cite{patzelt2017nonlinear}, we will investigate in detail how accurately the nonlinear master curves and rescaling exponents described above can be reproduced using propagator models \cite{bouchaud2004fluctuations} and their generalisations \cite{eisler2012price,taranto2016linear}. Our results provide important constraints for such models, and for realistic market models in general, since they quantify how the market reacts to both order bias and price-change probability on all intra-day time scales. Naturally, such improved impact models are of interest both for practitioners trying to reduce their trading costs and for regulators trying to understand the stability of markets.



\FloatBarrier 

\begin{acknowledgments}
This work was inspired by a preliminary unpublished study by Julius Bonart. We thank him, and M. Benzaquen, G. Bormetti, F. Bucci, J. De Lataillade, J. Donier, Z. Eisler, M. Gould, S. Gualdi, S. Hardiman, J. Kockelkoren, F. Lillo, I. Mastromatteo, D. Taranto \& B. Toth for many inspiring discussions. We also thank G. Bolton, J. Lafaye, and C.A. Lehalle for support and advice during data preparation.
\end{acknowledgments}

\bibliography{patzelt2017impact}

\begin{thebibliography}{10}

\bibitem{nobel2013background}
KVA.
\newblock {Economic Sciences Prize Committee of the Royal Swedish Academy of
  Sciences. Understanding Asset Prices}.
\newblock Nobelprize.org. Nobel Media AB 2013.
  \url{http://www.nobelprize.org/nobel_prizes/economic-sciences/laureates/2013/advanced.html},
  Oct. 14 2013.

\bibitem{farmer2009equilibrium}
J.~Doyne Farmer and John Geanakoplos.
\newblock The virtues and vices of equilibrium and the future of financial
  economics.
\newblock {\em Complexity}, 14(3):11--38, 2009.

\bibitem{lyons2000microstructure}
Richard~K. Lyons.
\newblock {\em The Microstructure Approach to Exchange Rates}, chapter~1.
\newblock MIT Press Cambridge, 2000.

\bibitem{shiller1981dividends}
Robert~J. Shiller.
\newblock Do stock prices move too much to be justified by subsequent changes
  in dividends?
\newblock {\em Am. Econ. Rev.}, 71:421--436, 1981.

\bibitem{cutler1989prices}
David~M. Cutler, James~M. Poterba, and H.~Summers, Lawrence.
\newblock What moves stock prices?
\newblock {\em Journal of Portfolio Management}, 15(3):4--12, 1989.

\bibitem{hopman2007prices}
Carl Hopman.
\newblock Do supply and demand drive stock prices?
\newblock {\em Quantitative Finance}, 7(1):37--53, 2007.

\bibitem{joulin2008news}
Armand Joulin, Augustin Lefevre, Daniel Grunberg, and Jean-Philippe Bouchaud.
\newblock Stock price jumps: news and volume play a minor role.
\newblock {\em Wilmott}, Sep. / Oct., 2008.

\bibitem{Bouchaud2009MarketsSlowlyDigest}
Jean-Philippe Bouchaud, J.~Doyne Farmer, and Fabrizio Lillo.
\newblock How markets slowly digest changes in supply and demand.
\newblock In Thorsten Hens and Klaus~Reiner Schenk-Hoppe, editors, {\em
  Handbook of Financial Markets: Dynamics and Evolution}. North-Holland,
  Elsevier, 2009.

\bibitem{donier2015fully}
Jonathan Donier, Julius Bonart, Iacopo Mastromatteo, and J-P Bouchaud.
\newblock A fully consistent, minimal model for non-linear market impact.
\newblock {\em Quantitative finance}, 15(7):1109--1121, 2015.

\bibitem{zarinelli2015beyond}
Elia Zarinelli, Michele Treccani, J.~Doyne Farmer, and Fabrizio Lillo.
\newblock Beyond the square root: Evidence for logarithmic dependence of market
  impact on size and participation rate.
\newblock {\em Market Microstructure and Liquidity}, 01(02):1550004, 2015.

\bibitem{kylei1985continuous}
Albert~S Kyle.
\newblock Continuous auctions and insider trading.
\newblock {\em Econometrica}, 53(6):1315--1335, 1985.

\bibitem{plerou2002quantifying}
Vasiliki Plerou, Parameswaran Gopikrishnan, Xavier Gabaix, and H~Eugene
  Stanley.
\newblock Quantifying stock-price response to demand fluctuations.
\newblock {\em Physical Review E}, 66(2):027104, 2002.

\bibitem{patzelt2017nonlinear}
Felix Patzelt and Jean-Philippe Bouchaud.
\newblock Nonlinear price impact from linear models.
\newblock {\em Preprint arXiv:1706.04163 [q-fin.TR]}, 2017.

\bibitem{mackintosh2016needIII}
Phil Mackintosh and Ka~Wo Chen.
\newblock The need for speed iii: Physics and a little trigonometry.
\newblock KCG Holdings, Inc.
  \url{https://www.kcg.com/news-perspectives/article/the-need-for-speed-iii-physics-and-a-little-trigonometry},
  2016.

\bibitem{mackintosh2016needV}
Phil Mackintosh and Ka~Wo Chen.
\newblock The need for speed v: How important is 1 ms?
\newblock KCG Holdings, Inc.
  \url{https://www.kcg.com/news-perspectives/article/the-need-for-speed-v-how-important-is-ms},
  5 2016.

\bibitem{dayri2015large}
Khalil Dayri and Mathieu Rosenbaum.
\newblock Large tick assets: implicit spread and optimal tick size.
\newblock {\em Market Microstructure and Liquidity}, 1(01):1550003, 2015.

\bibitem{lillo2003econophysics}
Fabrizio Lillo, J~Doyne Farmer, and Rosario~N Mantegna.
\newblock Econophysics: Master curve for price-impact function.
\newblock {\em Nature}, 421(6919):129--130, 2003.

\bibitem{jones1994transactions}
Charles~M Jones, Gautam Kaul, and Marc~L Lipson.
\newblock Transactions, volume, and volatility.
\newblock {\em Review of financial studies}, 7(4):631--651, 1994.

\bibitem{farmer2004really}
J.~Doyne Farmer, L{\`a}zl{\`o} Gillemot, and Fabrizio Lillo.
\newblock What really causes large price changes?
\newblock {\em Quantitative Finance}, 4(4):383--397, 2004.

\bibitem{gomber2015liquidity}
Peter Gomber, Uwe Schweickert, and Erik Theissen.
\newblock Liquidity dynamics in an electronic open limit order book: An event
  study approach.
\newblock {\em European Financial Management}, 21(1):52--78, 2015.

\bibitem{bouchaud2006random}
Jean-Philippe Bouchaud, Julien Kockelkoren, and Marc Potters.
\newblock Random walks, liquidity molasses and critical response in financial
  markets.
\newblock {\em Quantitative finance}, 6(02):115--123, 2006.

\bibitem{madhavan1997security}
Ananth Madhavan, Matthew Richardson, and Mark Roomans.
\newblock Why do security prices change? a transaction-level analysis of nyse
  stocks.
\newblock {\em Review of Financial Studies}, 10(4):1035--1064, 1997.

\bibitem{taranto2014adaptive}
DE~Taranto, G~Bormetti, and F~Lillo.
\newblock The adaptive nature of liquidity taking in limit order books.
\newblock {\em Journal of Statistical Mechanics}, 2014(6), 2014.

\bibitem{lillo2004long}
Fabrizio Lillo, J~Doyne Farmer, et~al.
\newblock The long memory of the efficient market.
\newblock {\em Studies in nonlinear dynamics \& econometrics}, 8(3):1, 2004.

\bibitem{gerig2008theory}
Austin Gerig.
\newblock {\em A theory for market impact: How order flow affects stock price}.
\newblock Phd thesis, arxiv:0804.3818, University of Illinois, 2007.

\bibitem{patzelt2013instability}
Felix Patzelt and Klaus Pawelzik.
\newblock An inherent instability of efficient markets.
\newblock {\em Sci. Rep.}, 3:2784, 2013.

\bibitem{patzelt2011critical}
Felix Patzelt and Klaus Pawelzik.
\newblock Criticality of adaptive control dynamics.
\newblock {\em Phys. Rev. Lett.}, 107:238103, Dec 2011.

\bibitem{fama1970efficientMarkets}
Eugene Fama.
\newblock Efficient capital markets: A review of theory and empirical work.
\newblock {\em The Journal of Finance}, 25(2):383--417, 1970.

\bibitem{bouchaud2004fluctuations}
Jean-Philippe Bouchaud, Yuval Gefen, Marc Potters, and Matthieu Wyart.
\newblock Fluctuations and response in financial markets: the subtle nature of
  `random'price changes.
\newblock {\em Quantitative finance}, 4(2):176--190, 2004.

\bibitem{eisler2012price}
Zoltan Eisler, Jean-Philippe Bouchaud, and Julien Kockelkoren.
\newblock The price impact of order book events: market orders, limit orders
  and cancellations.
\newblock {\em Quantitative Finance}, 12(9):1395--1419, 2012.

\bibitem{taranto2016linear}
Damian~Eduardo Taranto, Giacomo Bormetti, Jean-Philippe Bouchaud, Fabrizio
  Lillo, and Bence Toth.
\newblock Linear models for the impact of order flow on prices i. propagators:
  Transient vs. history dependent impact.
\newblock {\em Preprint arXiv:1602.02735 [q-fin.TR]}, 2016.

\end{thebibliography}

\appendix
\section*{Appendices}

\section{Single trade impacts}
\label{sec:single_trade_impact}

\begin{figure*}[p]
  \centering
  \includegraphics[width=\textwidth]{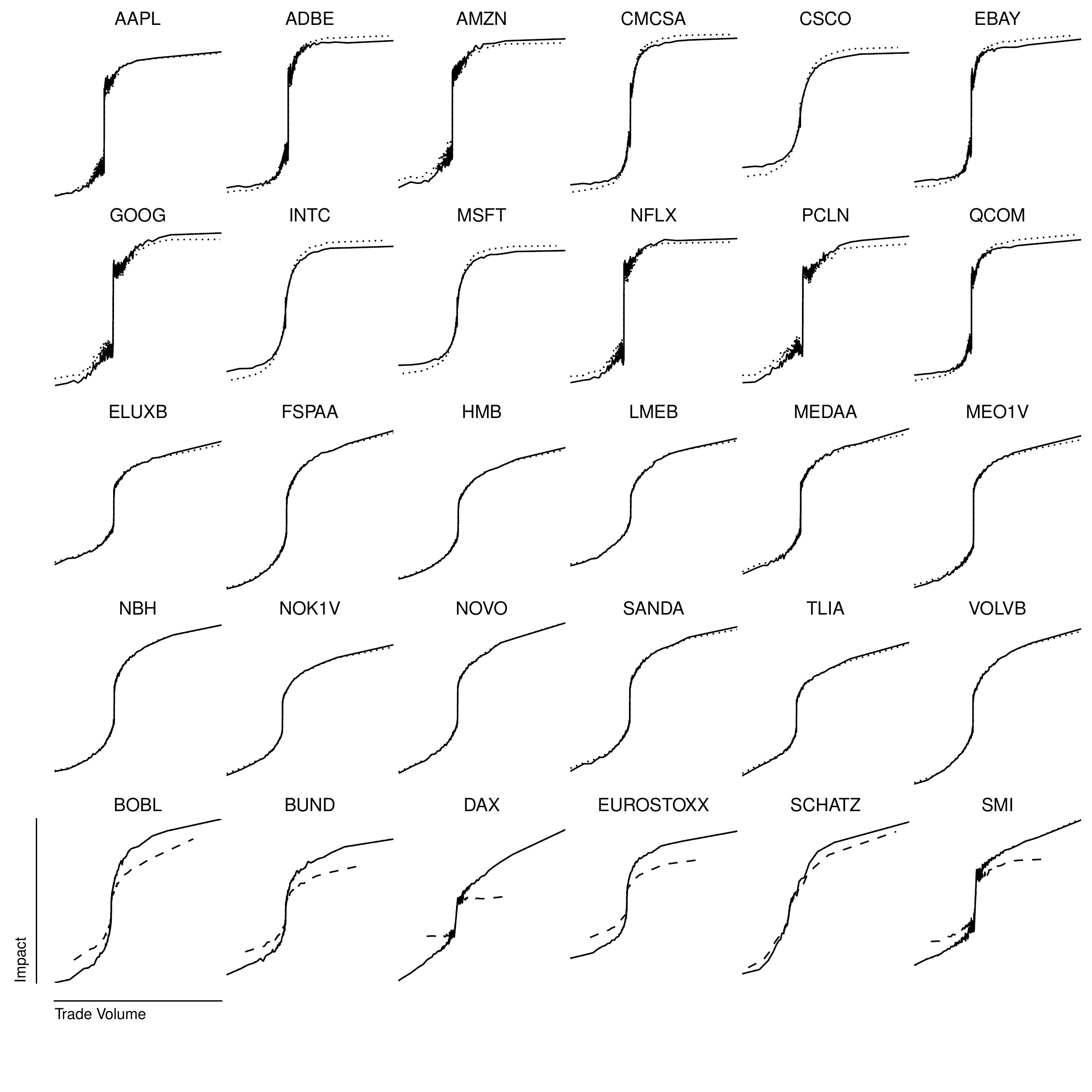}
  \caption{Single-trade impacts as a function of the order volume imbalance. First two rows: NASDAQ stocks, 2012-2016. Rows three and four: OMX stocks, 2012-2016. Lowest row: EUREX futures (10/2014-12/2015). Solid lines: estimated trade-sign. Dotted lines: exchange-provided trade-signs  (not available for hidden liquidity on NASDAQ). Dashed lines: exchange-provided sign and trade-id (only EUREX).}
  \label{fig:rdV-1-trade-impacts-all}
\end{figure*}

Fig.~\ref{fig:rdV-1-trade-impacts-all} shows mid-price-impacts for single (estimated) trades for different instruments. The impacts are dominated by a step at the origin and a highly concave dependency on volume. In other words: The one-trade impact strongly depends on the direction of the trade and much less on its volume. In practice, it is challenging to measure the true one-trade impact because it is not known which transactions belong to the same trade. As shown in the lowest row, merging transactions with the same signs and timestamps (solid lines) aggregates more trades than merging based on the trade id reported by the exchange (dashed lines). Unfortunately, since most available data does not contain trade-ids, one usually can only calculate an $N$-trade impact with a small $N > 1$.  Since we consider aggregate trades for $N \gtrsim 10$ in the following, however, this slight underestimation of the true sizes of the aggregation bins should only lead to minor quantitative differences. Note that the impacts could be measured using the transaction price instead of the mid-price. In this case, the one-trade impact is dominated by the bid-ask-bounce and the aggregate impact quickly converges towards the one for the mid-price (not shown).

\section{Impact curves for 30 instruments}
\label{sec:all_impacts}

\begin{figure*}[p]
  \centering
  \includegraphics[width=\textwidth]{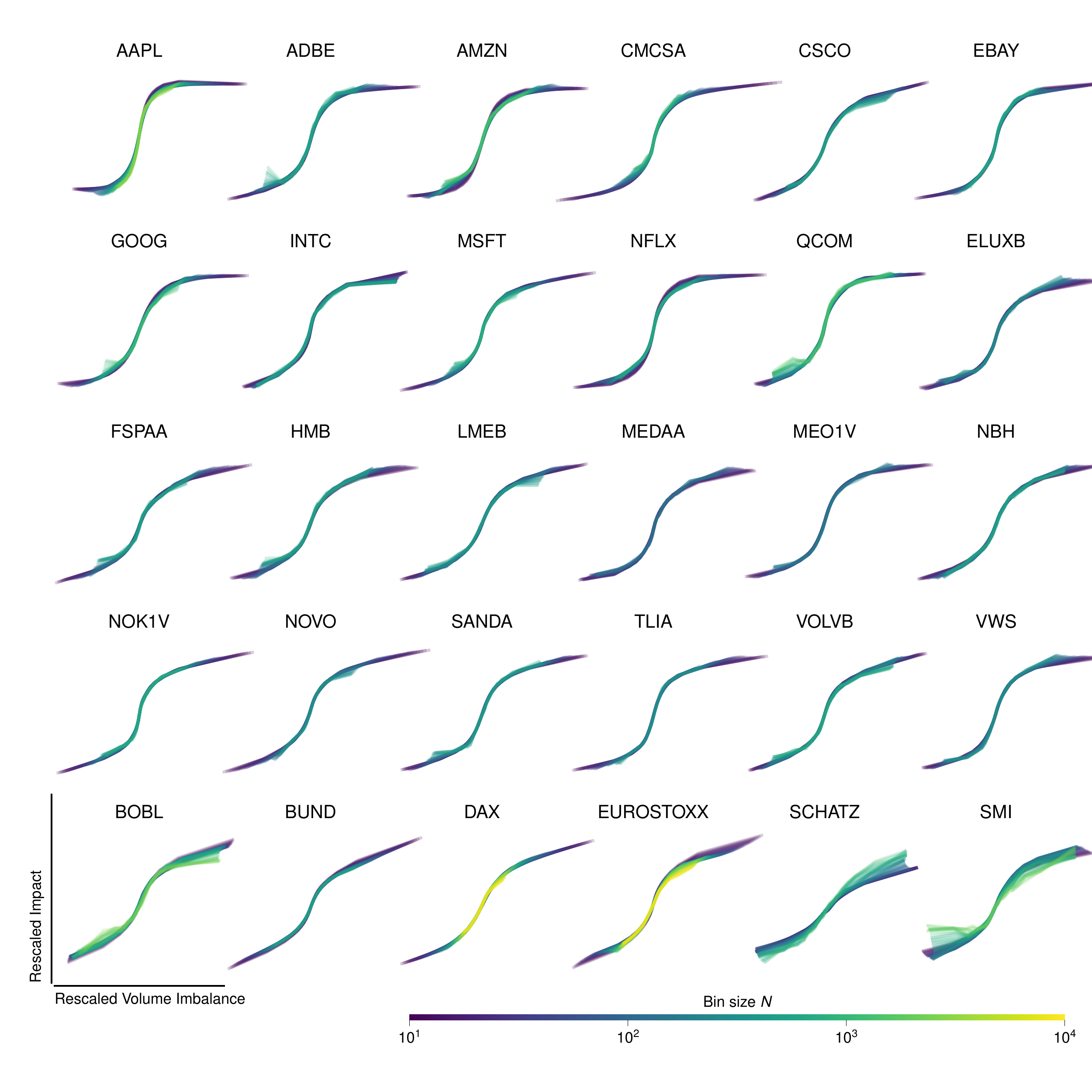}
  \caption{Price impact as a function of the order volume imbalance. The raw point-clouds were quantile-binned along the imbalance axis. Therefore the curves have a constant noise level but a range of imbalances that changes with the temporal bin size $N$. Instruments and years as in Fig.~\ref{fig:rdV-1-trade-impacts-all}.}
  \label{fig:rdV-rescaled-impacts-all}
\end{figure*}

Fig.~\ref{fig:rdV-rescaled-impacts-all} shows aggregate order-volume-impacts for different instruments, for different $N$, and rescaled according to Eq.~\ref{eq:scaling_form}.
Aggregate impact is never fully linear and exhibits a universal shape on all intra-day scales of each instrument with small difference across instruments. Most notably, small tick instruments (like AAPL and GOOG) tend to have a saturating impact or even a small overshoot while large-tick instruments (like MSFT, the nordic stocks and the futures), do not fully saturate for large imbalances.

\begin{figure*}[p]
  \centering
  \includegraphics[width=\textwidth]{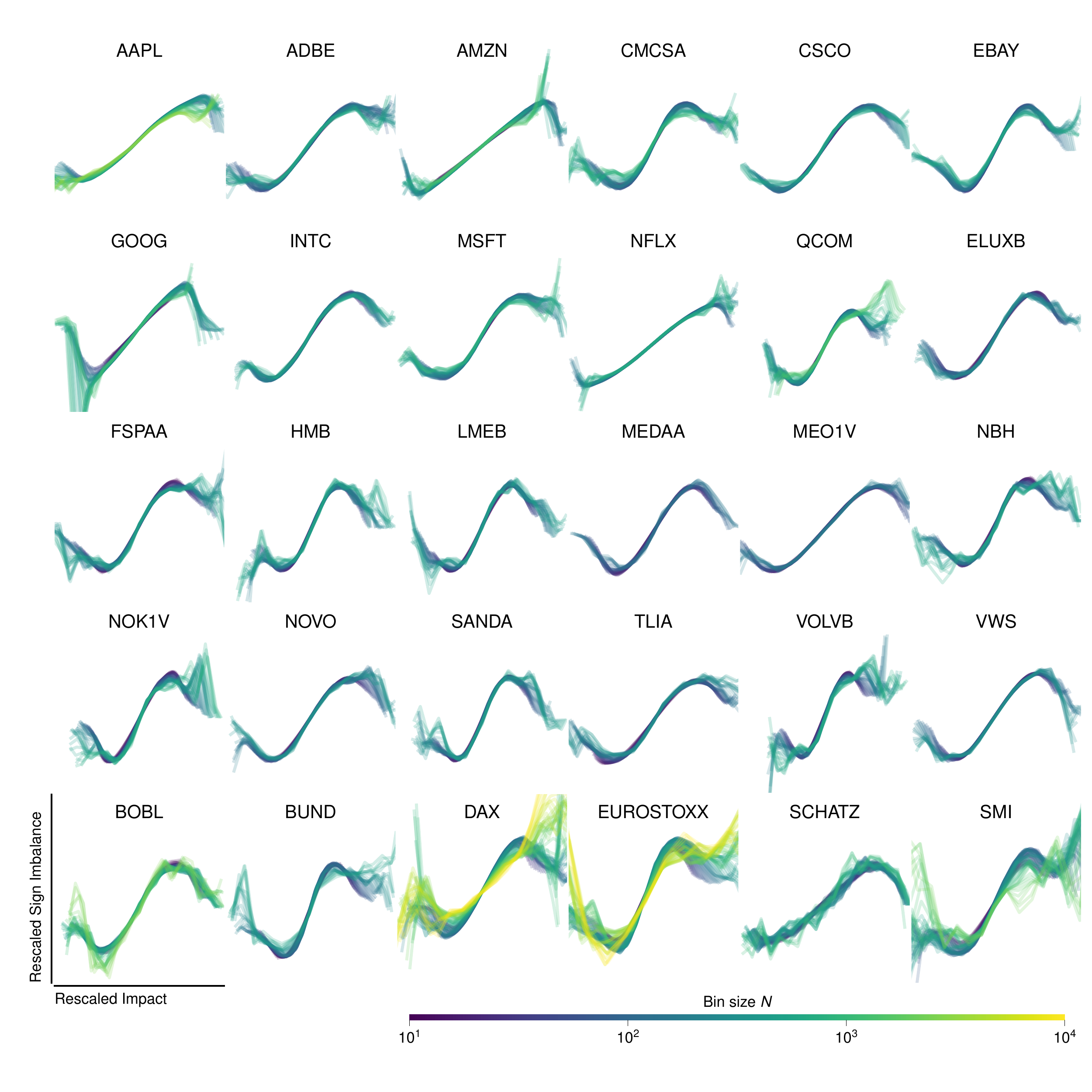}
  \caption{Price impact as a function of the order sign imbalance. Instruments and years as in Fig.~\ref{fig:rdV-1-trade-impacts-all}. }
  \label{fig:rdS-rescaled-trade-impacts-all}
\end{figure*}

\begin{figure*}[p]
  \centering
  \includegraphics[width=\textwidth]{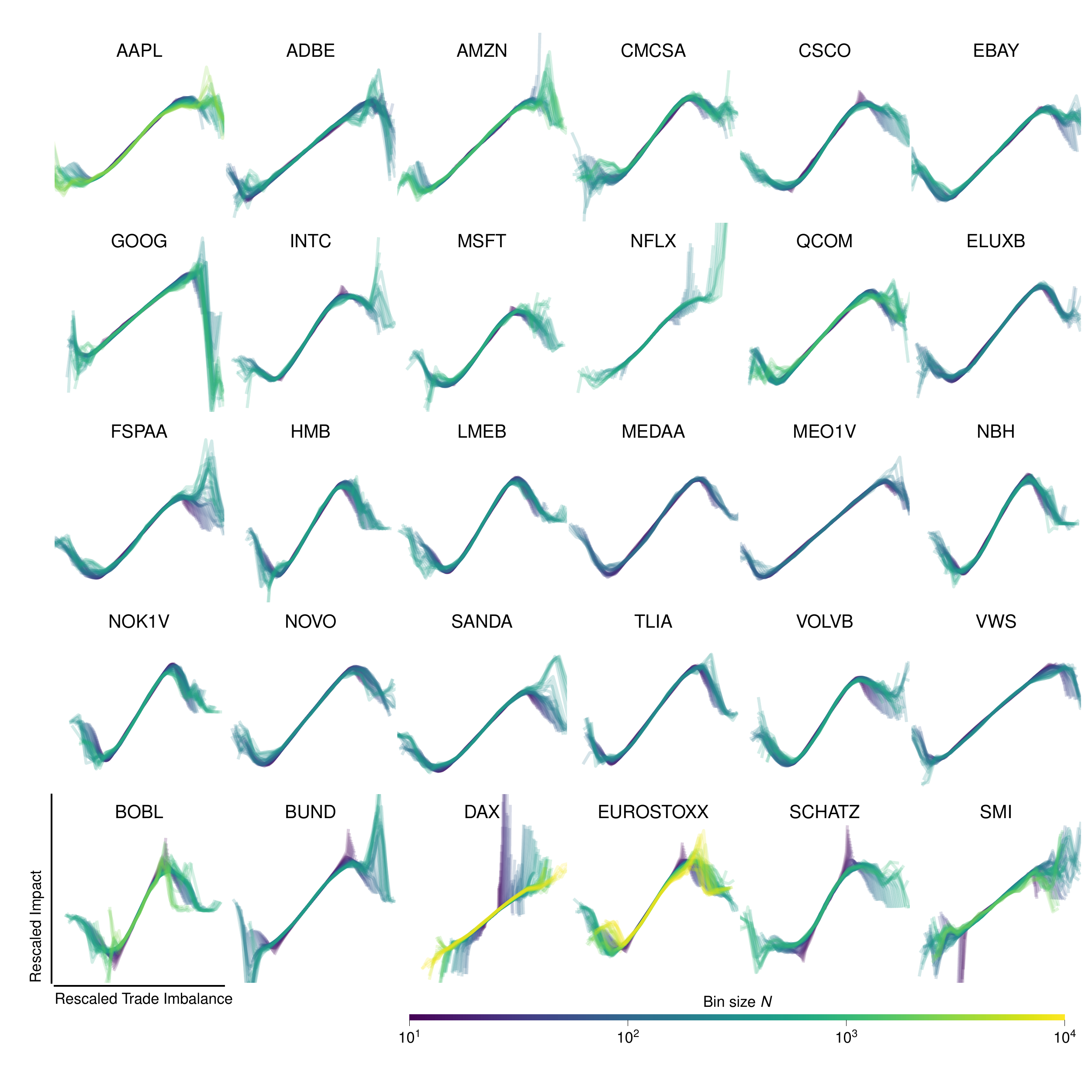}
  \caption{Price impact as a function of the trade imbalance. Instruments and years as in Fig.~\ref{fig:rdV-1-trade-impacts-all}.}
  \label{fig:rdTi-rescaled-trade-impacts-all}
\end{figure*}

Fig.~\ref{fig:rdS-rescaled-trade-impacts-all} shows aggregate order-sign-impacts for for different instruments. Similar results are found for the trade-imbalance in Figure~\ref{fig:rdS-rescaled-trade-impacts-all}. The only exception are the DAX futures, which seem to exhibit reverting impacts conditioned on large order-sign imbalances but not for large trade imbalances.

\section{Change probability curves for 30 instruments}
\label{sec:change-all}

\begin{figure*}[p]
  \centering
  \includegraphics[width=\textwidth]{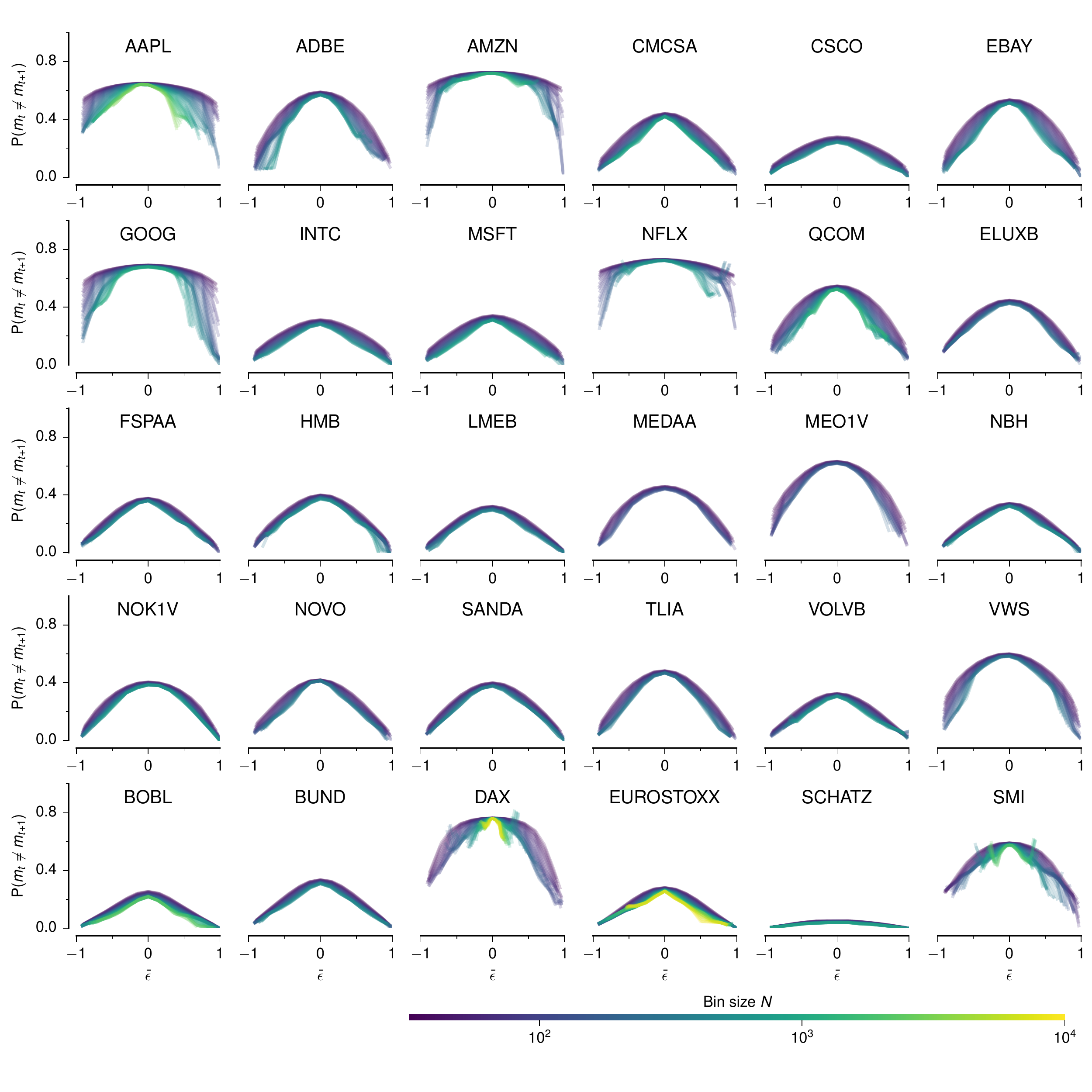}
  \caption{Probability (within a bin) that the mid-price changes between two trades vs the mean sign. Instruments and years as in Fig.~\ref{fig:rdV-1-trade-impacts-all}.}
  \label{fig:change-all}
\end{figure*}

Fig.~\ref{fig:change-all} shows the price-change probability for different instruments. To emphasize the approximate invariance of the curve over a wide range of bin sizes for all but the most small-tick instruments, we here set the lowest bin-size to 32.

\end{document}